\documentclass[onecolumn,superscriptaddress,nopacs,english,pre,longbibliography,notitlepage]{revtex4}
\usepackage{booktabs}
\usepackage{tabularx} 
\usepackage{amsmath}  
\usepackage{graphicx} 
\usepackage[margin=1in,letterpaper]{geometry} 
\usepackage[final]{hyperref} 
\usepackage{amssymb}
\usepackage{mathrsfs}
\usepackage{multirow}
\usepackage{longtable}
\usepackage{threeparttable}
\usepackage{amsthm}

\usepackage{algorithm} 
\usepackage{algpseudocode}  
\usepackage{amsmath}  

\usepackage{float}
\newtheorem{theorem}{Theorem}

\hypersetup{
	colorlinks=true,       
	linkcolor=blue,        
	citecolor=blue,        
	filecolor=magenta,     
	urlcolor=blue         
}

\begin{document}

\title{Shortcut Matrix Product States and its applications}

\author{Zhuan Li$^{12}$, and Pan Zhang}

\email{panzhang@itp.ac.cn}
\affiliation{
CAS key laboratory of theoretical physics, Institute of Theoretical Physics, Chinese Academy of Sciences, Beijing 100190, China.\\
$^{2}$Physics department, University of Chinese Academy of Science.
}


\begin{abstract} 
Matrix Product States (MPS), also known as Tensor Train (TT) decomposition in mathematics, has been proposed originally for describing an (especially one-dimensional) quantum system, and recently has found applications in various applications such as compressing high-dimensional data, supervised kernel linear classifier, and unsupervised generative modeling. However, when applied to systems which are not defined on one-dimensional lattices, a serious drawback of the MPS is the exponential decay of the correlations, which limits its power in capturing long-range dependences among variables in the system. To alleviate this problem, we propose to introduce long-range interactions, which act as shortcuts, to MPS, resulting in a new model \textit{ Shortcut Matrix Product States} (SMPS).
When chosen properly, the shortcuts can decrease significantly the correlation length of the MPS, while preserving the computational efficiency. 
We develop efficient training methods of SMPS for various tasks, establish some of their mathematical properties, and show how to find a good location to add shortcuts. 
Finally, using extensive numerical experiments we evaluate its performance in a variety of applications, including function fitting, partition function calculation of $2-$d Ising model, and unsupervised generative modeling of handwritten digits, to illustrate its advantages over vanilla matrix product states. 
\end{abstract}

\maketitle

\section{Introduction}
Tensor network (TN) is a powerful tool in representing and analyzing systems defined in an high dimension. In quantum many-body physics, tensor networks appear as a natural tool for representing quantum many-body states using a moderate number of parameters~\cite{jiang2008accurate,huckle2013computations,huang2018generalized}, that is for compressing the big Hilbert space by limiting bond dimensions, while preserving relevant physics~\cite{orus2014practical}. 
Recently years, tensor networks have drawn many attentions in diverse fields out of physics. In applied mathematics, tensor networks are known as tensor decompositions~\cite{oseledets2011tensor,cichocki2018tensor,zhao2016tensor,kolda2009tensor,cichocki2015tensor}. In machine learning, researchers found tensor networks very useful as a kernel linear classifier, in the field of generative modeling, in analyzing neural networks, as well as in natural language processing~\cite{cohen2017analysis,stoudenmire2016supervised,PhysRevX.8.031012,gallego2017physical,pestun2017tensor}.
The most famous tensor network is probably the Matrix Product States (MPS), which consists of series of $3$-way tensors. Also known as Density Matrix Renormalization Group (DMRG)\cite{white1992sr,schollwock2005density}, MPS has become a standard method for dealing with one dimensional quantum systems with hopefully area law of entanglement entropy.
From the view point of computation, the $1$-D nature of MPS, together with canonical form make it very straightforward and efficient to calculate the contractions, expectation, and the normalization factors\cite{orus2014practical,PhysRevX.8.031012}. This is actually one advantage of MPS against many other powerful tensor networks. For a simple example, PEPS \cite{orus2014practical} is a more natural choice than MPS for $2$-D models and data such as natural images. However, contracting PEPS is a very hard problem, thus we need to adopt approximations\cite{TRGLevin,xie2012coarse}. This is opposed to MPS where we have polynomial algorithms for contracting a MPS exactly.

In applied mathematics, the MPS is known as \textit{Tensor Train} (TT) format\cite{oseledets2011tensor} in the field of tensor decomposition which asks to represent a high dimensional tensor as a multilinear product of small tensors with a moderate number of totally parameters. It has been found that while having asymptotically the same number of parameters as canonical decompositions such as Canoinical Polyadic (CP) and Tucker decompositions \cite{kolda2009tensor}, MPS enjoys much stable decomposition algorithms based on low rank approximations via singular value decompositions. Recently, MPS has been employed to machine learning tasks. In \cite{stoudenmire2016supervised}, authors proposed to use MPS in the supervised learning for natural images and showed that the MPS achieves a performance close to the state-of-the-art methods in the MNIST dataset. In this task, MPS works as a linear classifier in the kernel space --- the Hilbert space given by mapping each pixel in the image to a spinor.
MPS has also been used in the unsupervised learning~\cite{PhysRevX.8.031012} as ``Born machine'' to represent squared root of joint probability distribution of variables in complex data. In ~\cite{novikov2015tensorizing} authors use MPS to compress fully connected layers in the convolution neural networks, archived a large compression rate and reduces heavily number of the total parameters while the representation power of the neural network is preserved.

Despite its wide applications in many fields, it is well known that MPS is finite-correlated\cite{orus2014practical,gallego2017physical}, which means that the two point correlation of the MPS decays exponentially with the distance of two tensors. 
Although the exponential decay of correlation does happen in many $1$-D physics systems (e.g. the ground state of gaped non-critical 1-D systems), due to this limitation MPS is not able to describe very well critical systems, and systems with long range dependences.
When applying MPS to other fields such as tensor decomposition and machine learning, the issue of exponential decay of correlation function become very serious. Because in these applications the data we care about, such as natural images, are usually not defined on $1$-D. Thus there are indeed long range correlations we really want to preserve when modeling the data using tensor networks. For example, when adopting MPS as a kernel linear classifier \cite{stoudenmire2016supervised} for images of size $L\times L$, one has to unfold the matrix of pixels to a vector. This unfolding operation increases distance of two nearby pixels in natural images from $1$ to $L$ in the MPS representations.

In this paper, we propose a new structure of MPS, namely \textit{Shortcut Matrix Product State} (SMPS), to ease the pain of the exponential correlation decay. We add shortcut to MPS which links a pair of distant tensors in the MPS, providing the extra correlations for the distant tensors. 
MPS with periodic condition can be viewed as a special case of SMPS where the first and the last tensors are linked. 
Recently, in \cite{zhao2016tensor} authors give some examples that when applied as high-way tensor representations, MPS with a periodic boundary condition (named as Tensor Ring by the authors) gives a better performance than the vanilla MPS.
Intuitively, due to the freedom in choosing the linked tensor, in principle SMPS can have a much better performance than MPS with periodic condition. 
Then a natural question arises is how to to predict the best location of the shortcuts, and how to develop efficient algorithm for training SMPS. On the shortcut location, our proposal is to define a correlation function $C(i,j)$ of each pair of tensors in the MPS, then by comparing the distribution of $C(i,j)$ between the training data and MPS, an efficient location for the the shortcut emerges. In this sense, SMPS also reveal some intrinsic properties of the training data because the shortcut in the MPS can be viewed as the contraction of the inner indices. On the training algorithms, although the shortcuts in the SMPS obviously breaks the canonical conditions in MPS, we develop two algorithms, one by decomposing an arbitrary tensor to SMPS by sequential singular value decompositions, and the other one being gradient-based learning algorithm using given loss function. Some mathematical properties are also established. We further show that analogous to MPS, there exists efficient direct sampling algorithm of SMPS since the normalization factor can be calculated quite easily by contracting all the tensor exactly.

We applied the SMPS to several problems, including compression of a high-order tensor representing a complex function and natural images, computing partition function of Ising models, and as a variational distribution to present joint distribution of pixels in natural images. Using extensive experiments we show that the shortcut in the MPS improve the performance in representing and dealing with high dimensional data, providing a more precise compression for image with less parameter, calculating a more accurate partition function for 2-D Ising model, and preserving better correlations in the unsupervised generating modeling.

The paper is organized as follows. In Section \ref{sec:ap}, we concisely review MPS, its applications and the issue of exponential decay of correlations. In Section \ref{sec:stt}, we present SMPS model, develop learning algorithms for several tasks. In Section \ref{sec:experiment}, we apply our model to highly oscillating function, real world image, 2-D Ising model and MNIST dataset. 

\section{Matrix product states, its  applications and the correlation decay issue}\label{sec:ap}

\begin{figure}[H]
    \centering
    \includegraphics[width=0.9\textwidth]{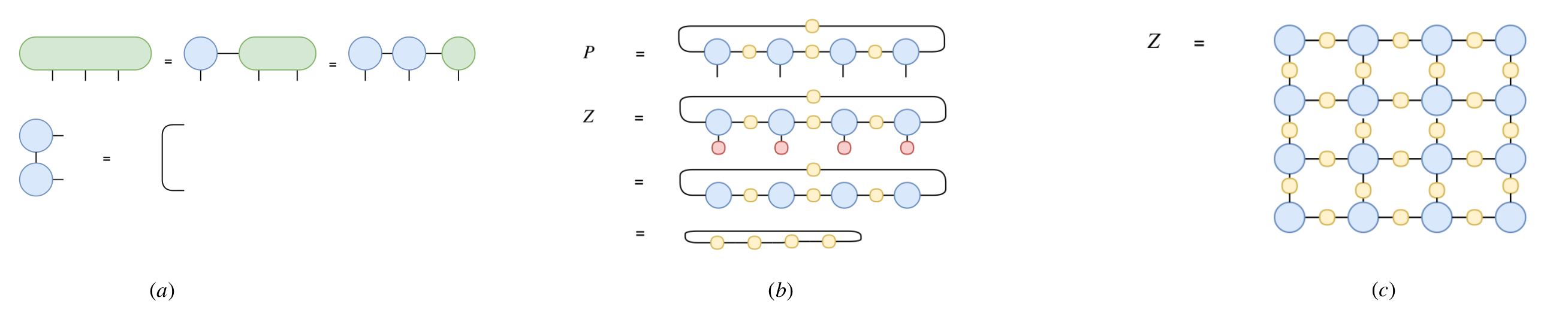}
    \caption{(a) Decomposing a tensor into MPS with canonical condition by employing sequential singular value decompositions. (b) Probability tensor and partition function of 1-D Ising model with periodic boundary condition. (c) Partition function of 2-D Ising model with open boundary condition.}
    \label{fig:notation1}
\end{figure}

Without loss of generality, let us consider a system composed of $N$ spin $1/2$, the dimension of the Hilbert space is $2^N$. The matrix product states represent quantum states of the system by interconnected 3-way tensors, each of which has shape $2D^2$ where $D$ stands for the bond dimension. Thus the total number of parameters of the MPS is asymptotically $2ND^2$, which is much smaller than $2^N$ when $D$ is not exponentially large.
There are ways to obtain the MPS. In the case that  we have a super computer to a wave function of the state $|\Psi\rangle$, or the given high way tensor , then we can split the system into two subsystems, as


\begin{equation}
|\Psi \rangle = \sum_{nm} \Psi_{nm} |a_n\rangle|b_m\rangle
\end{equation}
where $\{|a_n\rangle, n = 1,2,\cdots, N\}$ and $\{|b_m\rangle, m = 1,2,\cdots, M\}$ are the two basis sets of these two subsystems, $|\Psi\rangle_{nm}$ is a  $N \times M$ matrix representation of the wave function. Then  Singular value decomposition (SVD) can be applied to decrease the number of parameters induced while preserve as much as possible the entanglement entropy.

\begin{equation}
\begin{split}
|\Psi\rangle & =\sum_{nm} \sum_{a=1}^{\min(N,M)} U_{na}S_{aa}V_{ma}^{\dagger}|a_n\rangle|b_m\rangle =\sum_{nm} \sum_a A^{(1)}_{na} A^{(2)}_{ma}|a_n\rangle|b_m\rangle 
\end{split}
\end{equation}
where $A^{(1)}_{na} = U_{na}$ and $A^{(2)}_{ma}=S_{aa}V_{ma}^{\dagger} $.
Therefore, we can obtain a MPS by using SVD repeatedly, which represents $|\Psi\rangle$ as $n$ interconnected $3$-way tensors. This is illustrated in Fig.~\ref{fig:notation1}(a)), and is known as Schmidt decomposition~\cite{orus2014practical}, or sequential SVD algorithm~\cite{oseledets2011tensor}. 
The MPS obtained in this way is known to be in the canonical forms where most of tensors, except one of them is isometric:
\begin{equation} \sum_{i,j}A^{(n)}_{ijk}A^{(n)}_{ijl}=\delta_{kl}, \quad \mathrm{if}\ n < N.
\end{equation}
Obviously if we set $\sum_{ijk}A^{(N)}_{ijk}A^{(N)}_{ijk}=1$, then the MPS can be also viewed as a normalized many-body states. 

In addition to representing quantum states, MPS have many more applications. Here we give several examples.
\paragraph{Curve fitting}
Besides the wave function of quantum many-body system, other high-order tensor data, which acts as natural generalization to data matrices, can be also (approximately) represented in MPS format for reducing number of parameters. 
For example, the vector containing values of a highly oscillating function or a matrix containing pixels of an image can be folded into a high-order tensor, and then represented in MPS with a low bond dimension but privides agood approximation.\cite{oseledets2011tensor, zhao2016tensor,khoromskij2015tensor,khoromskij2015fast}. The method to transform the high-order tensor into the MPS format is called MPS or Tensor Train decomposition~\cite{oseledets2011tensor} which minimizes the Frobenius norm between the high-order tensor and MPS with given maximum bond dimension $d_{max}$. Compared with CP decomposition and Tucker decompositions, MPS decompostion yeilds asymptotically the same number of parameteres, but but having a much more stable decomposition algorithm ~\cite{oseledets2011tensor}. 

\paragraph{Partition function of classic statistical mechanics system}
Another important application of tensor networks is computing partition function or free energy of a classic statistical mechanics problem.
We first briefly review the Ising model in tensor diagram notation. Consider a $d$-dimensional lattice formed by a set of adjacent lattice sites. The spin of each site is represented by a discrete variable $\sigma_i\in \{-1,+1\}$. Any two lattice sites have an interactions $J_{ij}$, and each lattice site has an external magnetic field $h_i$. The energy of each configuration $\sigma$ is given by
\begin{equation}
H(\sigma)=-\sum_{i,j}J_{ij}\sigma_i\sigma_j - \sum_ih_i\sigma_i
\end{equation}
The configuration probability is given by the Boltzmann distribution:
\begin{equation}
    \begin{split}
        \mathbb{P}(\sigma)=\frac{e^{-\beta H(\sigma)}}{Z}, \quad Z=\sum_\sigma e^{-\beta H(\sigma)}
    \end{split}
\end{equation}
For the 1-D Ising model where $J_{ij}=J, h_i=h$, the probability tensor $P$ and the partition function $Z$ is straightforward, see Fig.~\ref{fig:notation1} (b).
In the figure, the blue tensors represent identity tensors, i.e., $T_{ijk}=\delta_{ij}\delta_{ik}$, and the yellow tensors are the Boltzmann factor matrices
$Y_{\sigma_i,\sigma_j} = e^{-\beta J_{ij}\sigma_i\sigma_j+\beta( \frac{1}{d_i}\sigma_ih_i+\frac{1}{d_j}\sigma_jh_j)},
$
and the red tensors denote the all-one tensor, $d_i$ denotes degree of node $i$ in the lattice.
The Boltzmann factor matrix $Y$ can be factorize as $Q\Lambda Q^{-1}$, where $\Lambda$ is the diagonal matrix whose diagonal elements are the corresponding eigenvalues and we have:
\begin{equation}
    Z=\mathbf{Tr}(Y^N)=\mathbf{Tr}((Q\Lambda Q^{-1})^N)=\mathbf{Tr}(\Lambda^N).
\end{equation}
So for the $1-$D model, at the thermodynamic limit with $N\to\infty$, we have $Z=\Lambda_{11}^N, N \to \infty$. This means in the $1$-D Ising model at the thermodynamic limit, the MPS with the maximum bond dimension $d_{max}=1$ can be a prefect approximation for the probability tensor.

For the 2-D Ising model, the exact partition function become more complex (see Fig.~\ref{fig:notation1} (c)). It can be naturally computed using contraction of a $2$-D tensor network called projected entangled pair states (PEPS). However, the exact contraction is $\#$ P-Hard \cite{schuch2007computational}. Therefore, some algorithms, such as tensor renormalization group (TRG) methods\cite{levin2007tensor,xie2012coarse, evenbly2015tensor} have been developed to approximately do the contraction. In this work we consider representing the probability tensor and compute the partition function of the $2-$D Ising model using MPS. 
There are two methods to obtain the MPS format of probability tensor for 2-D Ising model. The first method is to minimize the free energy of samples generated from the MPS, and the second way is to apply the MPS decomposition to the $2-$D probability tensor network for the Ising model, then by contracting the MPS with the all-one tensor, we can obtain the partition function of the the $2-$D Ising model.

\paragraph{Unsupervised and supervised learning}
MPS has been recently employed for supervised learning. In the supervised learning, every training sample $\tau$ has a label $\tau^l$. By adding the extra index into the MPS, one obtains a label $\psi(\tau)^l$ after contract the MOS and training element. Then the quadratic loss function can be defined~\cite{stoudenmire2016supervised}:
\begin{equation}
    L=\frac{1}{2}\sum_{\tau \in T}\sum_l(\psi(\tau)^l-\tau^l)^2,
\end{equation}
and consists of adjusting of parameters of MPS for decreasing the loss function.
Similar to unsupervised learning, DMRG and two-site update algorithm are used in minimizing $L$. It has been shown in~\cite{stoudenmire2016supervised} that MPS trained in this way achieves a performance close to the state-of-the-art classifiers on the MNIST dataset.

When the data come with no labels, and the unsupervised learning task is to represent the joint distribution of data, MPS can be used as a generative model, in the context of Born Machine \cite{PhysRevX.8.031012}. In this setting, we view MPS as the wave function, and the probability amplitude $\psi(\tau)$ is obtained by contracting MPS for each data. Therefore, we can use MPS to model the joint probability distribution of the training data set, by minimizing the negative log-likelihood(NLL) cost function~\cite{PhysRevX.8.031012}:
\begin{equation}
    L=-\frac{1}{N}\sum_{\tau\in T}\mathrm{log}(\mathbb{P}(\tau))=-\frac{1}{N}\sum_{\tau\in T}\mathrm{log}\left(\frac{|\psi(\tau)|^2}{Z}\right)
\end{equation}
where $T$ is the set of training samples, and $N$ is the size of training set. 

By using density-matrix renormalization group(DMRG)\cite{white1992sr} and two-site update algorithm, one can minimize $L$ which is equivalent to minimizing KL divergence~\cite{kullback1951} between two the empirical data distribution and model distribution\cite{PhysRevX.8.031012,kullback1951information}. Due to the canonical condition, the partition function of the MPS can be calculated exactly and straightforward. We have:
Moreover, by using the zipper algorithm, MPS can generate samples unbiasly and directly.\cite{PhysRevX.8.031012,ferris2012perfect}. 

Despite the wide applications of MPS, it is known that the correlation function of an MPS decreases exponentially as distance increases\cite{orus2014practical,gallego2017physical}. This was shown easily by using e.g eignvalues of transfer matrix~\cite{orus2014practical}. Here to illustrate this, we take a simple example by looking at the correlation function defined as (also see \cite{orus2014practical}).
\begin{equation}
    R(r)=\langle O_iO_{i+r}'\rangle-\langle O_i\rangle\langle O_{i+r}'\rangle
\end{equation}
where $O_i$ and $O_{i+r}$ are one body operators at site $i$ and $i+r$. 
Consider a very simple operator that $O_i$ and $O_{i+r}'$ are two diagonal matrix with diagonal elements $O_1, O_2$ and $O_3, O_4$ respectively. We have:
\begin{align}
        &R(r)=\langle O_iO_{i+r}'\rangle-\langle O_i\rangle\langle O_{i+r}'\rangle\nonumber\\
        &=\left(\mathbb{P}(11)O_1O_3+\mathbb{P}(12)O_1O_4+\mathbb{P}(21)O_2O_3+\mathbb{P}(22)O_2O_4\right)-(\mathbb{P}_i(1)O_1+\mathbb{P}_{i}(2)O_2)(\mathbb{P}_{i+r}(1)O_3+\mathbb{P}_{i+r}(2)O_4)\nonumber\\
        &=\mathrm{Tr}\left(\left[               
  \begin{array}{ccc}   
    \mathbb{P}(11) & \mathbb{P}(12) \nonumber\\  
    \mathbb{P}(21) & \mathbb{P}(22)\nonumber\\  
  \end{array}
\right] 
\left[                 
  \begin{array}{ccc}   
    O_1O_3 & O_2O_3 \nonumber\\  
    O_1O_4 & O_2O_4\nonumber\\  
  \end{array}
\right]\right)-   \mathrm{Tr}\left(\left[               
  \begin{array}{ccc}   
    \mathbb{P}_i(1)\mathbb{P}_{i+r}(1) & \mathbb{P}_i(1)\mathbb{P}_{i+r}(2) \nonumber\\  
    \mathbb{P}_i(2)\mathbb{P}_{i+r}(1) & \mathbb{P}_i(2)\mathbb{P}_{i+r}(2)\nonumber\\  
  \end{array}
\right] 
\left[                
  \begin{array}{ccc}   
    O_1O_3 & O_2O_3 \nonumber\\  
    O_1O_4 & O_2O_4\nonumber\\  
  \end{array}
\right]\right) \\
&=\mathrm{Tr}\left(\left(\left[               
  \begin{array}{ccc}   
    \mathbb{P}(11) & \mathbb{P}(12) \\  
    \mathbb{P}(21) & \mathbb{P}(22)\\  
  \end{array}
\right] -\left[              
  \begin{array}{ccc}   
    \mathbb{P}_i(1)\mathbb{P}_{i+r}(1) & \mathbb{P}_i(1)\mathbb{P}_{i+r}(2) \\  
    \mathbb{P}_i(2)\mathbb{P}_{i+r}(1) & \mathbb{P}_i(2)\mathbb{P}_{i+r}(2)\\  
  \end{array}
\right] \right) 
\left[                 
  \begin{array}{ccc}   
    O_1O_3 & O_2O_3 \nonumber\\  
    O_1O_4 & O_2O_4\\  
  \end{array}\nonumber,
\right]\right)
\end{align}
where $\mathbb{P}(xy)$ is the joint probability i.e. probability of $(\sigma_i=x)\wedge(\sigma_{i+r}=y)$. 

Let $A=\left[               
  \begin{array}{ccc}   
    \mathbb{P}(11) & \mathbb{P}(12) \\  
    \mathbb{P}(21) & \mathbb{P}(22)\\  
  \end{array}
\right]$ and $\lambda_1$ and $\lambda_2$ are two singular values of $A$ and $\lambda_1>\lambda_2$. We have:
\begin{equation}
\begin{split}
    R(r)&=\mathrm{Tr}\left(\left(A-A\left[                \begin{array}{ccc}   
    1 & 1 \\  
    1 & 1\\  
  \end{array}
\right]A\right)\left[                 
  \begin{array}{ccc}   
    O_1O_3 & O_2O_3 \\  
    O_1O_4 & O_2O_4\\  
  \end{array}
\right]\right) \\
&= \mathrm{det}(A) \left( O_1-O_2\right)(O_3-O_4)
\end{split}
\end{equation}
Where $|\mathrm{det}(A)|=\lambda_1\lambda_2$. Because $\mathbb{P}(11)+\mathbb{P}(12)+\mathbb{P}(21)+\mathbb{P}(22)=1$, we have:
\begin{equation}
    \frac{1-\sqrt{1-2|\mathrm{det}(A)|}}{2}\leq \lambda_2 \leq |\mathrm{det}(A)|
\end{equation}
where $C(i,i+r)=\lambda_2$ is the correlation function we define.
Since the correlation $R(r)$ decreases exponentially with the increase of distance $r$, the correlation function $C(i,i+r)$ also decreases exponentially with $r$ increases.

\begin{figure}[H]
    \centering
    \includegraphics[width=0.9\textwidth]{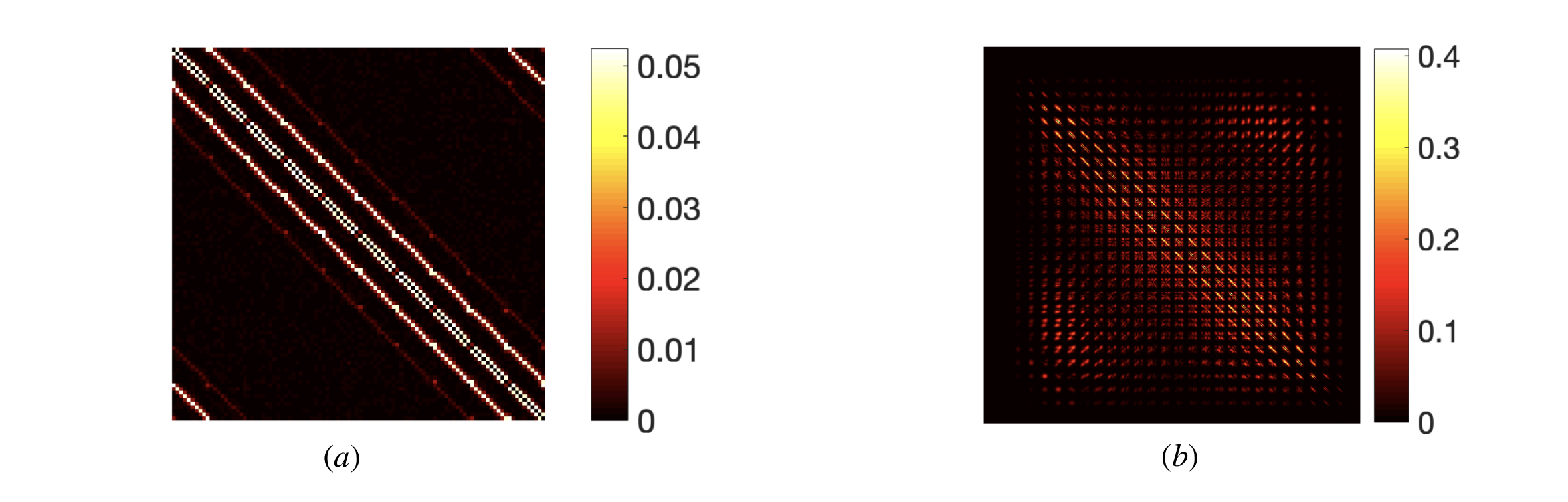}
    \caption{Distributions of correlation functions in different system.
    (a) 2-D Ising model with periodic boundary condition. There are four maximums each line, represent the four neighbor spins for each site. The correlations near the top-right and bottom-left corners also represent the periodic conditions in this case. (b) 10000 MNIST images. The correlations concentrate on the diagonal represent the short range interactions. The correlations near the top-right corner represent the long range interactions between the top and bottom part of image. The whole distribution divided into several sub-distribution, and the neighbouring sub-distribution are similar probably due to the shift invariance. 
    } 
    \label{fig:correlations}
\end{figure}

This problem could be very serious when MPS is used for tasks that long range dependences need to be captured. For examples, in Fig.~\ref{fig:correlations} we plot distributions of correlation functions in $2-D$ Ising model, as well as in the hand-written digits of MNIST datset. Where we can see that there are lots of strong correlation among pair of variables with a very long distances, for example on the off-diagonal positions of figures. 

\section{Shortcut Matrix Product States and the training algorithms}\label{sec:stt}

The exponential decay of correlations in MPS comes from the fact that the correlation between distance pairs of sites need to be built up by crossing many sites in between.
For MPS of $n$ sites with open boundary condition, to build up correlations between site $i$ and site $i+r$, one needs to cross $r$ sites, with $1\leq r\leq n$.
A natural idea for building up directly the correlations at key sites that are far away would be adding some shortcuts to decrease effective the distance between all pairs of sites. Actually the periodic boundary condition is a special case of this by adding short-cuts on the head and tail of the MPS. Then the distance one needs to cross for building correlations between site $i$ and site $i+r$ becomes $\min (r,n-r)$, which could be much smaller than $r$. The MPS with periodic boundary condition has been explored recently in~\cite{zhao2016tensor} for compression high-order tensors, where authors have shown that in synthetic and real-world datasets, the MPS with periodic boundary (called \textit{Tensor Ring} by the authors) condition is more expressive than MPS with open boundary condition.

In this section we will validate the idea of adding shortcuts on other locations of MPS. The first question we would like to address is where to add shortcuts. In this work we propose to use a very simple idea by testing the correlation functions. 

Assume we have a $T=T_{i_1i_2i_3\cdots},$ where $ i_1, i_2, i_3 \cdots \in \{1,2\}$. To describe the interaction strength of each pair of spin in these system, we define interaction matrix $Q^{(nm)}$ and correlation function $C(n,m)$. 

Without loss of generality, let us consider the interaction matrix for a probability tensor $P$, such as probability distribution of training dataset and probability tensor $P$ of Ising model, is defined as:
\begin{equation}
    Q^{(nm)}_{i_1,i_2}=P(\sigma_{n}=i_1, \sigma_{m}=i_2)
\end{equation}
The interaction matrix for a wave function $\Psi$ is defined as:
\begin{equation}
    \includegraphics[width=0.4\textwidth]{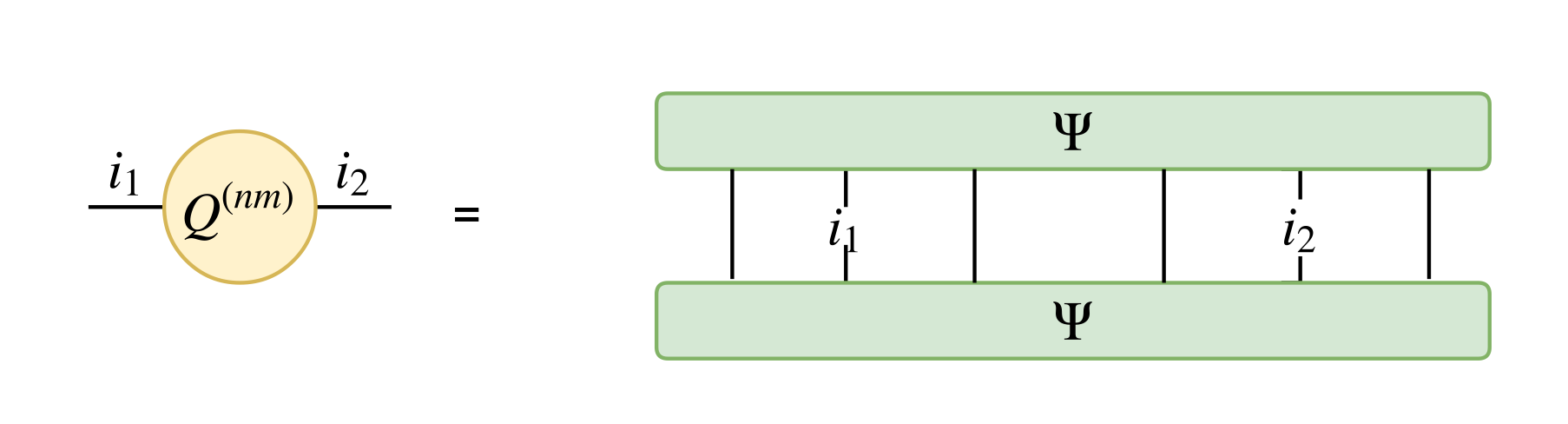}
\end{equation}

The correlation function $C(n,m)$ is defined as the second singular value of the interaction matrix. The reason why the second singular value is chosen is that MPS with $d_{max} = 1$ is good enough to store all of the first singular values. Therefore, MPS is enough for the simple system such as 1-D infinite Ising model. However, the correlation function of some systems could have complex distributions (see Fig.~\ref{fig:correlations} ). Equipped with the correlation function, we can compare correlation distributions of the real data, such as training images, and the distribution generated by the MPS, calculate the differences and choose the location that the difference is maximized.

Based on giving position of the shortcut, in the following part we develop two algorithms for obtaining the SMPS given the shortcut position. The first one is to decompose an arbitrary high-order tensor to SMPS using singular value decompositions, and the second algorithm is a learning algorithm similar to DMRG, which gradually adjust parameters of a SMPS using gradients with respect to the loss function, together with truncating the singular value spectrum for limiting number of parameters.

\subsection{SMPS decomposition of an arbitrary tensor}
\begin{figure}[H]
    \centering
    \includegraphics[width=0.8\textwidth]{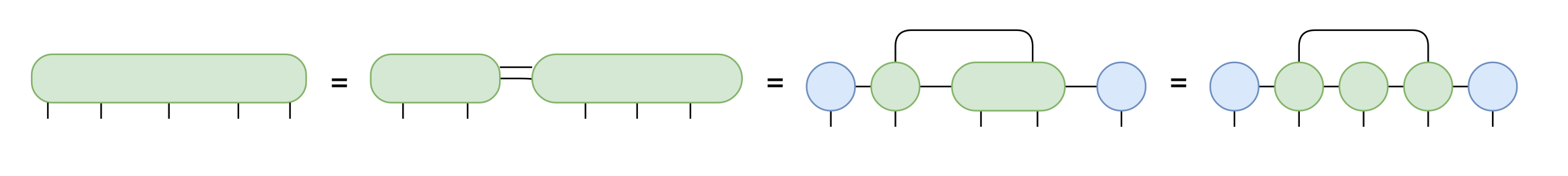}
    \caption{Decomposing a tensor into SMPS by employing singular value decomposition. It is clear that a part of canonical conditions are broken.}
    \label{fig:decompose}
\end{figure}

Let $T$ be a $N$-th order tensor of size $w_1 \times w_2 \times \cdots \times w_N $. 
The unfolding matrix $T_k$ of tensor $T$ is defined by unfolding (reshaping) the tensor in to a matrix:
\begin{equation}
\begin{split}
T_k&=T({w_1\cdots w_k}; {w_{k+1}\cdots w_n})=\mathrm{reshape}(T, \prod ^k_{s=1}w_s,  \prod ^n_{s=k+1}w_s),
\end{split}
\end{equation}
The SVD is applied on this unfolding matrix, and then the result can be written as the contraction of two tensors:
\begin{equation}
    T_k=U_kS_kV_k'=\sum_{a}T^{(1)}_{w_1,w_2,\cdots,w_k,a}T^{(2)}_{a,w_{k+1},\cdots,w_n}.
\end{equation}
Tensor $T^{(1)}$ and $T^{(2)}$ are further decomposed into two tensors using the same procedure. By repeating this process, tensor $T$ can be transformed into contraction of a sequence of tensors $A^{(k)},k=1,2,\cdots , N $. In this way, we get the (non-shortcut) MPS . To get the shortcuts at the certain location in tensor train, we split the bond during the first SVD process, and then permute the bond to the predetermined positions. For example, we have:
\begin{equation}
\begin{split}
    T_k=U_kS_kV_k^\top&=\sum_{a_1}T^{(1)}_{w_1,w_2,\cdots,w_k,a_1}T^{(2)}_{a_1,w_{k+1},\cdots,w_N}\\
    &=\sum_{l_1,l2}T^{(1)}_{w_1,w_2,\cdots,w_k,l_1,l_2}T^{(2)}_{l_1,l_2,w_{k+1},\cdots,w_N}\\
    &=\sum_{l_1,l2}T^{(1)}_{w_1,\cdots,w_s,l_1,\cdots,w_k,l_2}T^{(2)}_{l_2,w_{k+1},\cdots,w_x,l_1,\cdots,w_N}\\
    &=\cdots=\sum_{a_i}A^{(1)}_{w_1,a_1}A^{(2)}_{a_1,w_2,a_2}\cdots A^{(s)}_{a_s,w_s,a_s,l_1}\cdots A^{(x)}_{a_x,w_x,a_x,l_1}\cdots A^{(N)}_{a_{n-1},w_N}
\end{split}
\label{equ:rank}
\end{equation}
where $l_1l_2=a$, and we link the $s$-th and $x$-th tensors. This gives the final SMPS. The whole process is depicted in Fig.~\ref{fig:decompose}.

During the sequence of SVDs, if there is no truncation on the singular spectrum carried out, that is when the decomposition is exact, we have the following property hold:
 \begin{theorem}
If the k-unfolding matrix $T_k$ has $rank (T_k) = R_k$, then there exists an SMPS decomposition with bound dimension $D_{k}$ and shortcut bound dimension $D_s$, which satisfied $D_sD_{k} \leq R_k$.
\end{theorem}
\begin{proof}
Consider the unfolding matrix $T_k$. Its rank is equal to $R_k$. After SVD we have:
\begin{equation}
\begin{split}
     T_k(w_1,\cdots, w_k; w_{k+1}, \cdots, w_n)&= \sum_{a=1}^{r_k} U(w_1,\cdots, w_k; a)S(a;a_1)V(a;w_{k+1}, \cdots, w_n)\\
     &=\sum_{a=1}^{r_k} U(w_1,\cdots, w_k; a)W(a;w_{k+1}, \cdots, w_n)
\end{split}
\end{equation}
where $r_k \leq R_k$.
Then we split the bond $a_1$ and permute new bonds into proper locations, we have:
\begin{equation}
    \begin{split}
        T_k(w_1,\cdots, w_k; w_{k+1}, \cdots, w_n)&= \sum_{l_1, l_2}^{D_s,D_{k+1}} U(w_1,\cdots, w_k; l_1, l_2)W(l_1, l_2;w_{k+1}, \cdots, w_n)\\
     &=\sum_{l_1, l_2}^{D_s,D_{k+1}} U(w_1,\cdots,l_1, \cdots , w_k; l_2)W(l_2 ;w_{k+1},\cdots, l_1, \cdots, w_n)
    \end{split}
\end{equation}
where $D_s*D_{k+1}=r_k$.
Now, matrix $U$ and $W$ can be viewed as the unfold matrix of tensor $\mathbf{U}$ and $\mathbf{W}$. After the decomposition, we have:
\begin{equation}
\begin{split}
    \mathbf{U}&=\sum_{a_i}A^{(1)}_{w1,a_1}\cdots A^{(s_1)}_{a_{s_1-1},l_1,a_{s_1}}\cdots A^{(k)}_{a_k,w_k,l_2}\\
    \mathbf{W}&=\sum_{a_i}A^{(k+1)}_{l_2,w_{k+1},a_{k+1}}\cdots A^{(s_2)}_{a_{s_2-1},l_1,a_{s_2}}\cdots A^{(n)}_{a_n,w_n}
\end{split}    
\end{equation}
Finally, we get the required SMPS:
\begin{equation}
    \mathbf{T}=\sum_{a_i,l_1,l_2}A^{(1)}_{w1,a_1}\cdots A^{(s_1-1)l_1}_{a_{s_1-2},w_{s_1-1},a_{s_1}}\cdots A^{(k)}_{a_k,w_k,l_2}A^{(k+1)}_{l_2,w_{k+1},a_{k+1}}\cdots A^{(s_2-1)l_1}_{a_{s_2-2},w_{s_2-1},a_{s_2}}\cdots A^{(n)}_{a_n,w_n}
\end{equation}
\end{proof}

The maximum bond dimension of the MPS generated from the method described above could still be exponentially large. Therefore, we need to set the maximum bond dimension $d_{max}$ in truncations. That is, during the SVD, only the first $d_{max}$ singular values are kept. Then we have:
\begin{equation}
    \begin{split}
        T_1&=U_1S_1V_1'=\sum_{a_1}T^{(1)}_{w_1,a_1}T^{(2)}_{a_1,w_{2},\cdots,w_n}=\sum A^{(1)}T^{(2)}_2\\
        &=\sum_{a_1}A^{(1)}_{w_1,a_1}U_2S_2V_2'=\sum_{a_1}A^{(1)}_{w_1,a_1}\sum_{a_2}T^{(2)}_{a_1,w_2,a_2}T^{(3)}_{a_3,w_{3},\cdots,w_n}\\
        &=\sum A^{(1)}\sum A^{(2)}T^{(3)}_2\\
        &=\cdots=\sum_{a_i}A^{(1)}_{w_1,a_1}A^{(2)}_{a_1,w_2,a_2}\cdots A^{(n)}_{a_{n-1},w_n}
    \end{split}
\end{equation}
At $i$-th step of SVD, the truncation can be carried out by controlling the truncation error as $\varepsilon_i^2$, then we have following theorem for controlling the total truncation error induced by the SMPS decomposition.
 \begin{theorem}
Assume at $i$-th step of singular value decomposition, truncation error is $\epsilon_i$, then the total $L_2$ distance between the original tensor $\mathbf{T}$ and SMPS decomposition is bounded above by ${\sqrt{\sum_i\varepsilon_i^2}}$.
\end{theorem}

 \begin{proof}
 At the first step, we have 
\begin{equation}
\begin{split}
     T_k(w_1,\cdots, w_k; w_{k+1}, \cdots, w_n)&=\sum_{a=1}^{r} U(w_1,\cdots, w_k; a)W(a;w_{k+1}, \cdots, w_n)\\
     \|\mathbf{T}-\mathbf{U}\otimes\mathbf{W}\|_2&=\sqrt{\varepsilon_1^2}
\end{split}
\end{equation}
Then we apply TT decomposition to tensor $\mathbf{W}$ and $\mathbf{U}$. According to \cite{oseledets2011tensor}, the $L_2$ distance $\Delta_2$ between tensor $\mathbf{W}$ and MPS-format $\mathcal{W}$, and the $L_2$ distance $\Delta_1$ between tensor $\mathbf{U}\otimes\mathcal{W}$ and MPS-format $\mathcal{U}\otimes\mathcal{W}$  are:
\begin{equation}
    \begin{split}
        \Delta_2&=\|\mathbf{W}-\mathcal{W}\|\leq \sqrt{\sum_{i_2}\varepsilon_{i_2}^2}\\
        \Delta_1&=\|\mathbf{U}\otimes\mathcal{W}-\mathcal{U}\otimes\mathcal{W}\|\leq \sqrt{\sum_{i_1}\varepsilon_{i_1}^2}
    \end{split}
\end{equation}
Therefore, the total $L_2$ loss between the original tensor and the final SMPS is:
\begin{equation}\label{err}
    \begin{split}
         \|\mathbf{T}-\mathcal{T}\|_2&\leq  \|\mathbf{T}-\mathcal{U}\otimes\mathcal{W}\|_2\leq  \|\mathbf{T}-\mathbf{U}\otimes\mathcal{W}\|_2+ \|\mathbf{U}\otimes\mathcal{W}-\mathcal{U}\otimes\mathcal{W}\|_2\\
         &\leq \|\mathbf{T}-\mathbf{U}\otimes\mathbf{W}\|_2+\|\mathbf{U}\otimes\mathcal{W}-\mathbf{U}\otimes\mathbf{W} \|_2+ \|\mathbf{U}\otimes\mathcal{W}-\mathcal{U}\otimes\mathcal{W}\|_2\\
         &\leq \sqrt{\sum_i\varepsilon_i^2}
    \end{split}
\end{equation}

\end{proof}

It is quite obvious that the SMPS can provide a better approximation than MPS because it keeps more singular values during the SVD for a given maximum bond dimension. The shortcut does not only add the number of parameters of the MPS, but also reveal inner structure of given tensor.
Interestingly, we can show that a SMPS can be seen as a sum of several MPSes, depending on the location of the shortcut. The relation is established in the following theorem.
\begin{theorem}\
A SMPS $T$ always has an equivalent expression as sum of multiple MPSes $T_s$, as $T=\sum_s T_s$.
\end{theorem}\label{sum}
\begin{proof}
Consider a SMPS $T$:
\begin{equation}
\begin{split}
T&=\sum_{i_1,\cdots,i_{N+1}} \sum_s A^{(1)\sigma_1}_{i_1i_2}A^{(2)\sigma_2}_{i_2i_3}\cdots A^{(l)\sigma_l}_{i_l i_{l+1}s}\cdots A^{(m)\sigma_m}_{i_mi_{m+1}s}\cdots A^{(N)\sigma_N}_{i_Ni_N+1}
\end{split}
\end{equation}
where $i_1=i_{N+1}=1$. Just interchange order of summation, we obtain:
\begin{equation}
\begin{split}
T&=\sum_s \sum_{i_1,\cdots,i_{N+1}} A^{(1)\sigma_1}_{i_1i_2}A^{(2)\sigma_2}_{i_2i_3}\cdots A^{(l)\sigma_l}_{i_l i_{l+1}s}\cdots A^{(m)\sigma_m}_{i_mi_{m+1}s}\cdots A^{(N)\sigma_N}_{i_Ni_N+1}=\sum_s T_s\\
\end{split}
\end{equation}
where $T_s$ are MPS.
\end{proof}

\subsection{Learning algorithm given a loss function}
\label{sec:loss}
In most of the practical applications, we are not able to store, or express the original tensor in the computer, then decompose it to a SMPS using the method developed in the last section.
One way to resolve the issue is using a learning algorithm to keep modifying a SMPS (starting from an initial one) to make it as close as possible to the target tensor by minimizing a proper loss function.
In this work we adopt two-site sweeping update for learning, where in each step of tensor update, we first contract (i.e. merge) two neighboring or linked (by shortcut) tensors into an higher order tensor $A$, then update their elements using gradient descent to decrease a loss function (such as $L_2$ loss or NLL loss), then use SVD and spectrum truncation to split the higher order tensor $T$ into two tensors to their original size. We repeat this process from the left most tensor to the right most tensor, then sweep back from right to left, until loss converges, or smaller than given criterion. 
The detailed process is given in Algorithm.\ref{alg:smps}:

\begin{algorithm}
	 \caption{SMPS Two-site Gradient Descent}  
	\label{alg:smps}  
	\leavevmode
  \begin{algorithmic}[1] 
    \Require  
      A $n$th-order tensor $T$ to approximate or a set of training samples $T$ to learn from, maximum bond dimension $d_{max}$, learning rate $\eta$, shortcut location $l_1, l_2$ and the prescribed relative error $\varepsilon$
    \Ensure  
      SMPS $\{A^{(i)}\}$ 
    \State Initialize random tensor $A^{(1)}\in \mathbb{R}^{1\times 2\times 2}$, $A^{(n)}\in \mathbb{R}^{2\times 2\times 1}$ and $A^{(i)} \in \mathbb{R}^{2\times 2\times 2}$ for $i=2,3,\dots,n-1$.
    \Repeat  
      \For {$i = 1 \ \mathrm{to} \ n-1$} \do \\
      \State Merge $A^{(i)}$ and $A^{(i+1)}$: $A \gets A^{(i)}\otimes A^{(i+1)} $.
      \State Compute the gradient tensor $\nabla_A L$ by using Eqn.\eqref{equ:L2} or Eqn.\eqref{equ:Ln}.  
      \State Update merged tensor $A$: $A \gets A + \eta \nabla_A L$. 
      \State Reshape merged tensor $A$ into matrix form
      \State Split merged matrix $M$ by applying truncated SVD with the rank $d \leq d_{max}$: $USV^t = \mathrm{svd}(M)$.
     \State Obtain $A^{(i)}$ by reshaping $U$ into proper size.
     \State Obtain $A^{(i)}$ by reshaping $SV^t$ into proper size.
     \If {$i = l_1$ or $i = l_2$}
     \State Merge $A^{(l_1)}$ and $A^{(l_2)}$: $A \gets A^{(l_1)}\otimes A^{(l_2)} $.
     \State Repeat step 5 to step 8.
     \State Obtain $A^{(l_1)}$ by reshaping $U$ into proper size.
     \State Obtain $A^{(l_2)}$ by reshaping $SV^t$ into proper size.
     \EndIf
     \State Obtain $L_2$ or $L_n$ loss. \EndFor
      \For {$i = n-1 \ \mathrm{to} \ 1$} \do \\
      \State Repeat step 4 to step 8
     \State Obtain $A^{(i)}$ by reshape $US$ into proper size.
     \State Obtain $A^{(i)}$ by reshape $V^t$ into proper size.
     \If {$i = l_1$ or $i = l_2$}
     \State Repeat step 12 and step 13.
     \State Obtain $A^{(l_1)}$ by reshape $US$ into proper size.
     \State Obtain $A^{(l_2)}$ by reshape $V^t$ into proper size. \EndIf
     \State Obtain $L_2$ or $L_n$ loss. \EndFor
    \Until{Relative change of loss is smaller than the given criterion or loss is smaller than $\varepsilon$ .}  
  \end{algorithmic} 
\end{algorithm}

In this work we consider three kinds of loss functions:
\paragraph {Contraction loss}
The contraction loss is simply the scalar obtained by contracting all the indices of the SMPS. 
As we have introduced in the Section \ref{sec:ap}, one application of tensor networks is calculating free energy of Ising models because the partition function $Z=\bigotimes_i A^{(i)}$ can be expressed as contraction of all virtual bonds of the tensor network. 
In this case, gradients of the loss function with respect to a tensor is simply written as
\begin{equation}
    \begin{split}
        \nabla_{A^{(k)}} Z = \bigotimes_{i \neq k} A^{(i)}
    \end{split}
\end{equation}
\paragraph{$L_2$ loss}
The (squared) $L_2$ loss between the SMPS, $\bigotimes_iA^{(i)} $, and the target tensor $T$ is defined as, 
   
\begin{equation}
\begin{split}
      L_2^2=\|T-\bigotimes_i A^{(i)}\|_2^2=\sum_{ijk\cdots}( T_{ijk\cdots}T_{ijk\cdots}+\bigotimes_iA^{(i)}_{ijk\cdots}\bigotimes_iA^{(i)}_{ijk\cdots}-2T_{ijk\cdots}\bigotimes_iA^{(i)}_{ijk\cdots})
\end{split}
\end{equation}

The gradient for the loss function with respect to a tensor $A$ is written as
\begin{equation}
\begin{split}
    &\nabla_A{ L_2^2}=2\frac{\partial \bigotimes_iA^{(i)}}{\partial A}(\bigotimes_iA^{(i)}-T)\\
    &\includegraphics[width=0.35\textwidth]{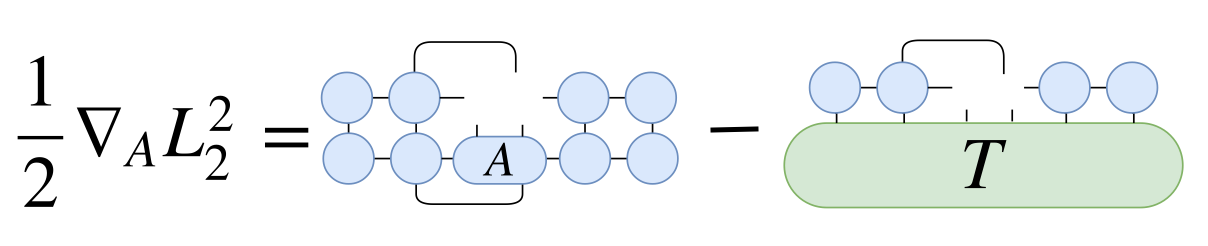}
\end{split}
\label{equ:L2}
\end{equation}
\paragraph{Negative log-likelihood (KL divergence) Loss}
For the applications of SMPS where the tensor represents a wave function, a joint distribution of observing a basis $\tau$ can be represented as \cite{PhysRevX.8.031012}
\begin{equation}
\mathbb{P}(\tau)) = \frac{1}{Z}\bigotimes_{i=1}^NA^{(i)}_{j_{i-1}\tau_ij_i\cdots}.
\end{equation}
When the target distribution is given by a set of data, as in Born machine, the loss function is the KL divergence between $\mathbb{P})$ and the data distribution $Q$
\begin{equation}
D_{\mathrm{KL}}=L_n-S(\mathrm{data}),
    \end{equation}
    where $S(\mathrm{data})$ denotes entropy of data distribution which is not a function of SMPS, and $L_n$ is the negative log likelihood loss function which is written as
\begin{equation}\label{eq1}
 L_n = -\frac{1}{N}\sum_{\tau \in T}{\mathrm{log}(\mathbb{P}(\tau))},
\end{equation}
where $T$ is the training set, $N$ is the size of training set. 
The gradients for the NLL loss is written as
\begin{equation}
\begin{split}
&\nabla_A{ L_n}=\frac{Z'}{Z}-\frac{2}{N}\sum_{\tau \in T}\frac{\Psi'(\tau)}{\Psi(\tau)}\\
&\includegraphics[width=0.4\textwidth]{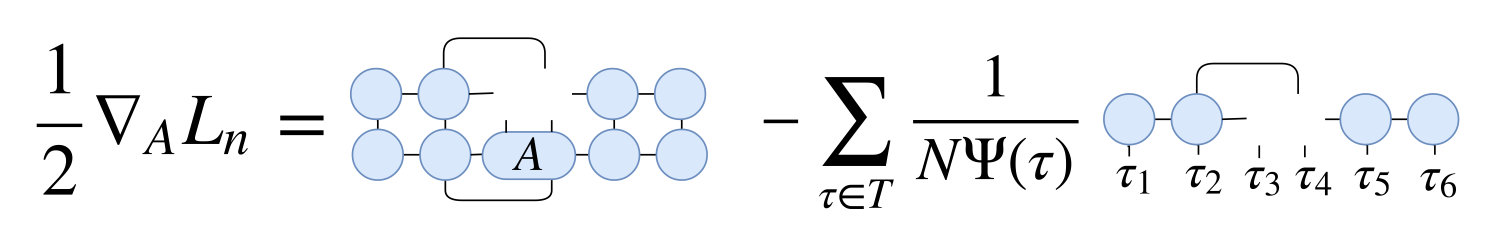}
\end{split}
\label{equ:Ln}
\end{equation}
where $\Psi'(\tau)$ is the derivative of the SMPS with respect to $A$, $Z'=2\sum_{\tau \in T}\Psi'(\tau)\Psi(\tau)$ and $\eta$ is the learning rate. Then, the two-site tensor is reshaped to a matrix. After the singular value decomposition, these two matrices are reshaped back to the tensors with proper orders. Even though SVD does not minimize the $L_n$ distance during decomposition, we can prove that it can provide a boundary for the the $L_n$ distance (see Sec.\ref{sec:apx}).
\section{Experimental results}\label{sec:experiment}
In this section we apply the SMPS and the training algorithms to several different applications by using different loss functions.
\subsection{Curve and image fitting}
The first application we consider is quite straightforward:
using SMPS to fit curves and images. 
Curves is a kind of $1$-D data, such as those produced by a one-variable functions, which can be expressed as a very long vector. An image can be expressed as a large matrix. These kinds of data can be tensorized into a multi-way tensor, then decomposed by SMPS for exploiting intrisic structures in the data for compressing.

\begin{figure}[H]
    \centering
    \includegraphics[width=0.9\textwidth]{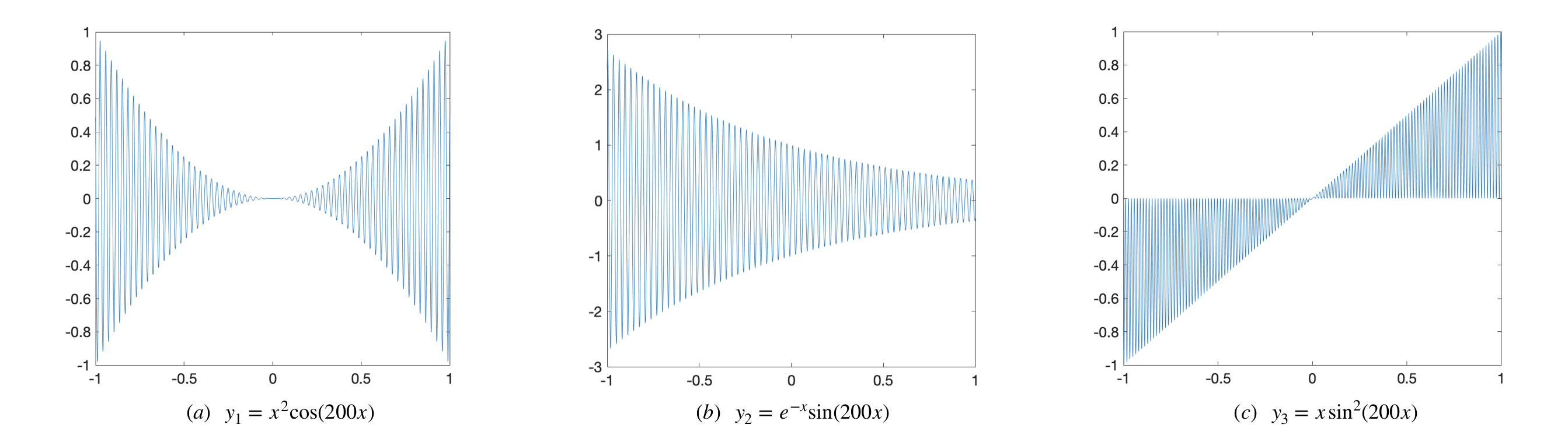}
    \caption{Highly oscillating functions}
    \label{fig:functions}
\end{figure}

Examples of curves are shown in Fig.~\ref{fig:functions}. The data given here are long vectors produced by descretization along the X-axis a highly oscillating functions. Apparently, the data is highly structured and is supposed to be compressed a lot against the straight-forward vector representation.
To do so, first we reshape the vector into $20$-th order tensors, and then decompose them into SMPS using learning algorithms developed in the last section with the $L_2$ loss between the original tensor and the obtained SMPS. 
Our results are listed in Table~\ref{Tab:curve1} where the number of parameters for MPS and SMPS are given for achieving a given fitting error $\epsilon$. We can see from the table that for all three function cures, to achieve the same amount of fitting error, SMPS and MPS requires similar number of parameters, and the number of parameters are much smaller than the original vector size $2^{20}$.
This is easy to understand because the oscillating functions contains a very strong pattern and does not really require long range dependences.

\begin{table}[]
\small
\centering
\begin{tabular}{|c|cc|cc|cc|}
\hline
\multirow{2}{*}{} & \multicolumn{2}{m{3.5cm}}{\centering{$y_1=x^2\mathrm{cos}(200x)$}} \vline&  \multicolumn{2}{m{3.5cm}}{\centering{$y_2=e^{-x}\mathrm{sin}(200x)$}}\vline& \multicolumn{2}{m{3.5cm}}{\centering$y_3=x\mathrm{sin}^2(200x)$} \vline\\
\cmidrule(r){2-3} \cmidrule(r){4-5} \cmidrule(r){6-7}
&  $\varepsilon$      &  $N$  
&  $\varepsilon$      &  $N$  
&  $\varepsilon$      &  $N$  \\
\hline
MPS &$1.81\times10^{-7}$   & 584   & $3.65\times10^{-13}$  & 152 & $3.01\times10^{-8}$& 762    \\
SMPS  &$1.81\times10^{-7}$   & 596   & $3.65\times10^{-13}$  & 172 & $3.01\times10^{-8}$& 762       \\
\hline
\end{tabular}
\caption{Results of curve fitting for highly oscillating functions in MPS/SMPS format with $20$ spins.
\label{Tab:curve1}}
\end{table}

\begin{table*}[]
\large
\small
\centering
\begin{tabular}{|c|cc|cc|cc|}
\hline
\multirow{2}{*}{} &\multicolumn{2}{m{3.5cm}}{\centering $y_1=x^2\mathrm{co s}(200x)+\mathcal{N}$} \vline& \multicolumn{2}{m{3.5cm}}{\centering$y_2=e^{-x}\mathrm{sin}(200x)+\mathcal{N}$}  \vline&\multicolumn{2}{m{3.5cm}}{\centering$y_3=x\mathrm{sin}^2(200x)+\mathcal{N}$} \vline\\
\cmidrule(r){2-3}  \cmidrule(r){4-5} \cmidrule(r){6-7}
&  $\varepsilon$      &  $N$  
&  $\varepsilon$      &  $N$  
&  $\varepsilon$      &  $N$  \\
\hline
MPS($d_{max}=30$)  &0.0453   & 20600   & 0.0149  & 20600 & 0.0401& 20600  \\
SMPS($d_{max}=20$) &0.0453   & 17960  & 0.0149  & 16360 & 0.0401& 16360  \\
\hline
\end{tabular}
\caption{Results of curve fitting for highly oscillating functions with noise in MPS/SMPS format with $20$ spins.
}
\label{Tab:curve2}
\end{table*}

To make the curve fitting more difficult and add long range dependences into the problem, we consider adding some random noise to the curve which apparently destroys the low-rank structure of the data. Thus both MPS and SMPS are expected to require more parameters than the noiseless case in curve fitting to achieve the same performance. Our results are shown in TABLE \ref{Tab:curve2}, where we can see that indeed the error one could achieve are heavily increased, as well as the number of parameters required. Here we can see that SMPS employs much less number of parameters than MPS in the noisy case where long-range dependence join the game.

\begin{figure}[H]
    \centering
    \includegraphics[width=\textwidth]{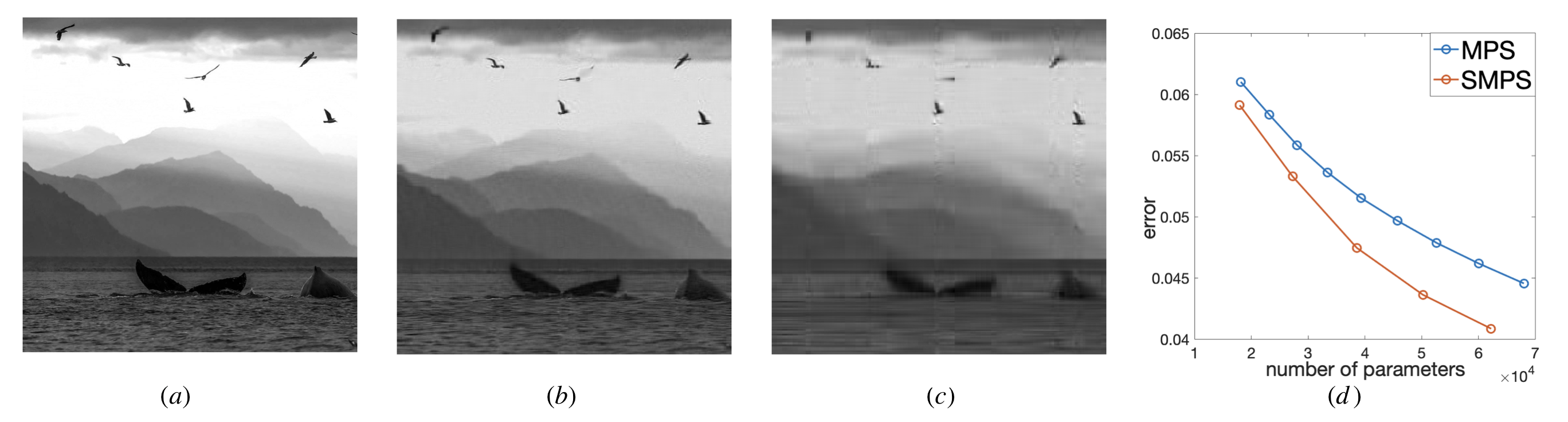}
    \caption{(a) $1024\times 1024$ image used in the test. (b) Image representation with error $\varepsilon = 0.044$. (c) Image representation with error $\varepsilon = 0.068$.(d) Result of the image representation.
    }
    \label{fig:image_result}
\end{figure}

Using MPS to compress matrices recently found applications in compressing deep neural networks, and has drawn lots of attention.
Nivikov et al~\cite{novikov2015tensorizing} showed that using MPS for compressing couplings matrix in the fully connected layers of the deep neural networks saves lots of parameters while only slightly decreases the performance of deep neural networks.
Then, we test our SMPS for image compression on real world images. Similar to the curve fitting, we reshape a $1024\times 1024$ image matrix into $20$-th order tensor, and then decompose it into SMPS. The image is more complex than the highly oscillating functions, therefore, more parameters in MPS and SMPS are required to represent the image tensor. The results are showed in Fig~\ref{fig:image_result} where we an see that SMPS with the same number of parameter has a much better performance than MPS, in the sense of giving much smaller fitting error.

The effect of shortcut can be easily understand in this application. During the decomposition, the small singular values are discarded after the truncated SVD. To minimize the error, the singular value should be kept as many as possible. A shortcut in MPS can indeed increase bond dimension hence keep more singular values. Therefore, the good locations for shortcut can also be determined using the distribution of the singular values.

\subsection{Partition function of the Ising model}
We can also transform probability tensor networks of 2-D Ising model showed in Section.\ref{sec:ap} into MPS/SMPS format. Then by contracting the MPS with the all-one tensor, we can compute the partition function $Z$ of the 2-D Ising model. From the partition function, we can get the energy $E$, magnetization $m$ and heat capacity $C$ of this model.

\begin{figure}[H]
    \centering
    \includegraphics[width=1\textwidth]{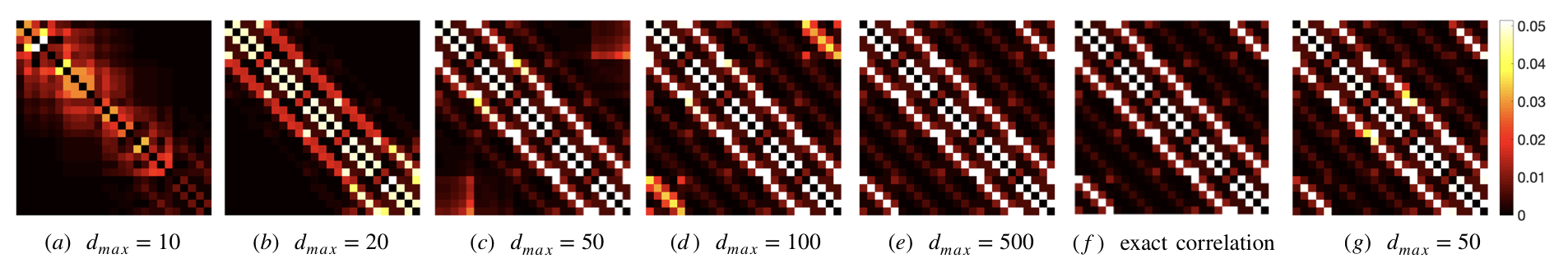}
    \caption{Distribution of correlation function of 2-D Ising model generated from MPS/SMPS. (a),(b),(c),(d) and (e) are generated from MPS. (f) is the exact correlation distribution. (g) is generated from SMPS, where the 5th and 23rd tensors are linked}
    \label{fig:link_ising}
\end{figure}

In transforming the tensor network into MPS/SMPS format, the maximum bond dimension $d_{max}$ for MPS are set to $10,20,50,100$ and $500$ respectively and $d_{max}$ for SMPS is set to $50$. The correlation generated from them are shown in Fig.~\ref{fig:link_ising}. We can see from the figure that MPS with a large bond dimension $d_{max}$ fits the exact solution very well (see Fig.~\ref{fig:link_ising}(f)). However with a small bond dimension, the distribution of the correlation function generated from MPS is far from the truth. As a comparison, we see that the SMPS with $d_{max}=50$ already captures very well the long range correlations as shown in the exact solution as shown in Fig.~\ref{fig:link_ising}$(f)$.  

Next, we calculate the partition function of this model by contracting all tenor with MPS/SMPS, and compare the results with $d_{max}=100$ to the exact solution~\cite{Onsager}. In addition to the free energy, energy $E=\frac{\partial Z}{\partial \beta}$ and heat capacity $\mathcal{C}=\frac{\partial E}{\partial T}$ are also calculated for comparison.  

\begin{figure}[H]
    \centering
    \includegraphics[width=1\textwidth]{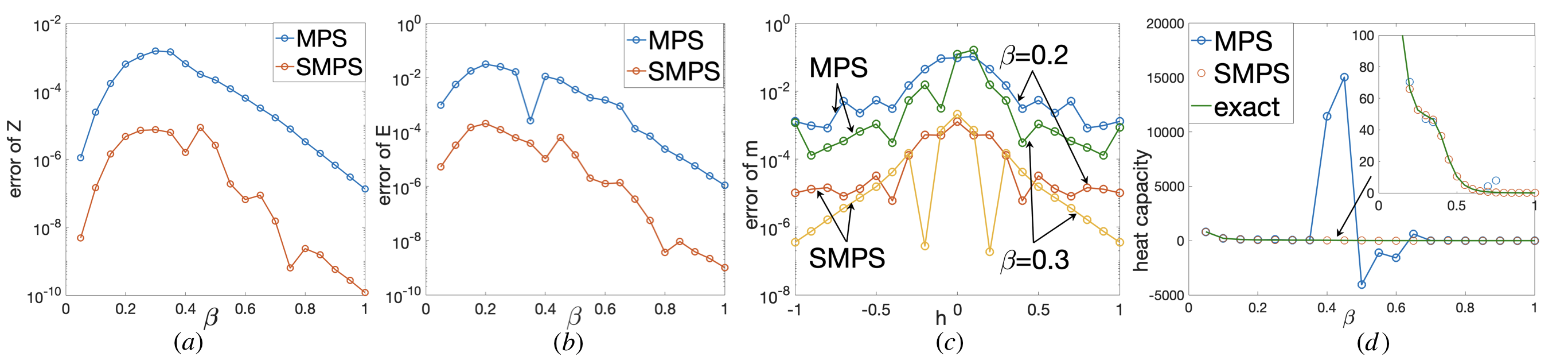}
    \caption{(a) Error of Z $=|Z_{100}-Z_{\textrm{exact}}|/Z_{\textrm{exact}}$.   (b) Error of E $=|E_{100}-E_{\textrm{exact}}|/E_{\textrm{exact}}$.  (c) Error of m $=|m_{100}-m_{\textrm{exact}}|/m_{\textrm{exact}}$.  (d) Heat capacity $=\frac{\mathrm{d} E}{\mathrm{d} T}=\frac{\mathrm{d} E}{\mathrm{d} \beta}\frac{\mathrm{d} \beta}{\mathrm{d} T}=-\frac{\mathrm{d} E}{\beta^2\mathrm{d} \beta}$}
    \label{fig:ising_result}
\end{figure}

Results are shown in Fig.~\ref{fig:ising_result}. We can see that the SMPS with the same bound dimension $d_{max}=100$ gives much more accurate results than MPS for all quantities we have tested. 
MPS with $d_{max}=100$ fails to compute the heat capacity $\mathcal{C}$ when $0.4<T<0.7$, even though it can gain the relatively accurate value of energy while SMPS with $d_{max}=100$ works perfectly for heat capacity in that regime. Obviously this indicates that the long range interaction plays a key rule. 


\subsection{Generative modeling for hand written digits}
In this section we use SMPS as a generative models for the MNIST dataset.
First, we vectorize the $28\times28$ gray scale images in MNIST dataset into vectors of size $784$. 

\begin{figure}[H] \includegraphics[width=1\textwidth]{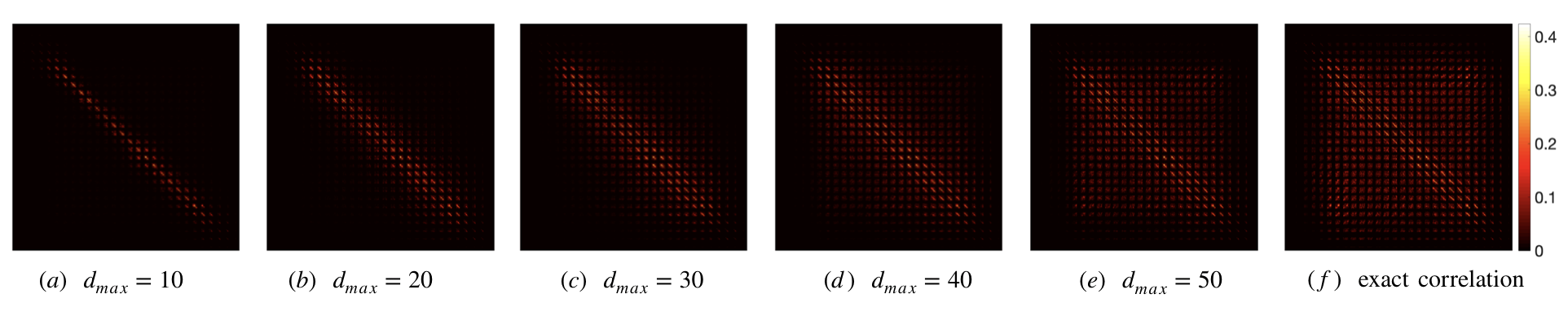}
\caption{The correlation distribution generated from MPS with different bond dimensions.
}
\label{fig:tt_mnist}
\end{figure}

To examine the MPS results, we do experiments on $50$ images randomly chosen from the MNIST dataset, train MPS with maximum bond dimension $d_{max}=10, 20, 30, 40, 50$ respectively. After 80 loops of training, the NLL of all MPS converge, the distribution of correlation function generated from these MPSes are showed in Fig.~\ref{fig:tt_mnist}.

The exact correlation computed are also plotted for comparison, where we can see that correlations concentrate on the nearby spins, while an atypical prominence occurs on the corner which implies the long range dependences in the images.
As expected, the span of correlations extend with the increase of the maximum bond dimension $d_{max}$ for MPS. When the maximum bond dimension becomes larger than the number of the images, the MPS can perfectly ``remember'' all of the training images, hence generate exact correlations. While MPS with small $d_{max}$ fail to represent long range correlations. 

\begin{figure}[htbp] \centering \includegraphics[width=1\textwidth]{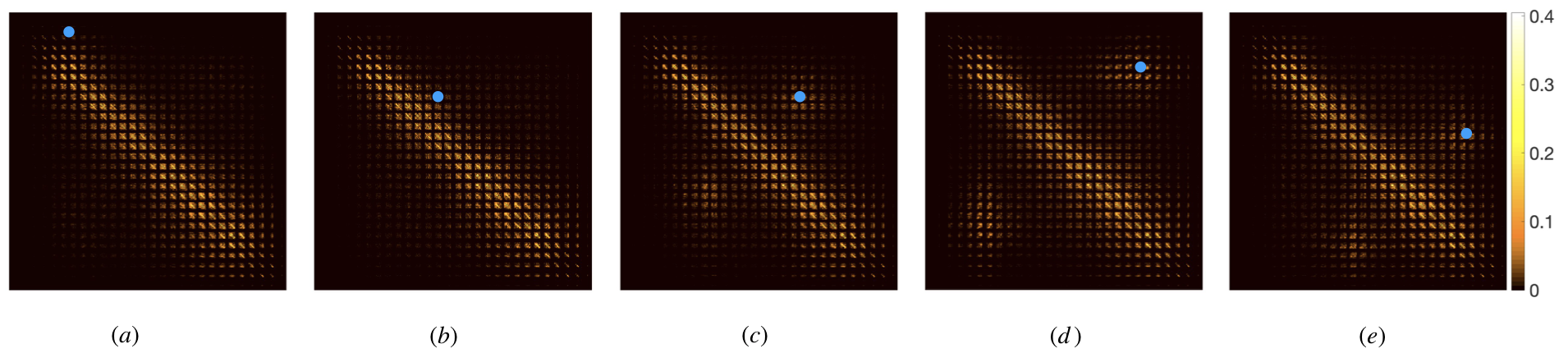}
\caption{The correlation distribution generated from SMPS with $d_{max}=20$, the blue dot denotes the location of the shortcut.
}
\label{fig:stt_mnist}
\end{figure}

\begin{figure}
    \centering
    \includegraphics[width=0.6\textwidth]{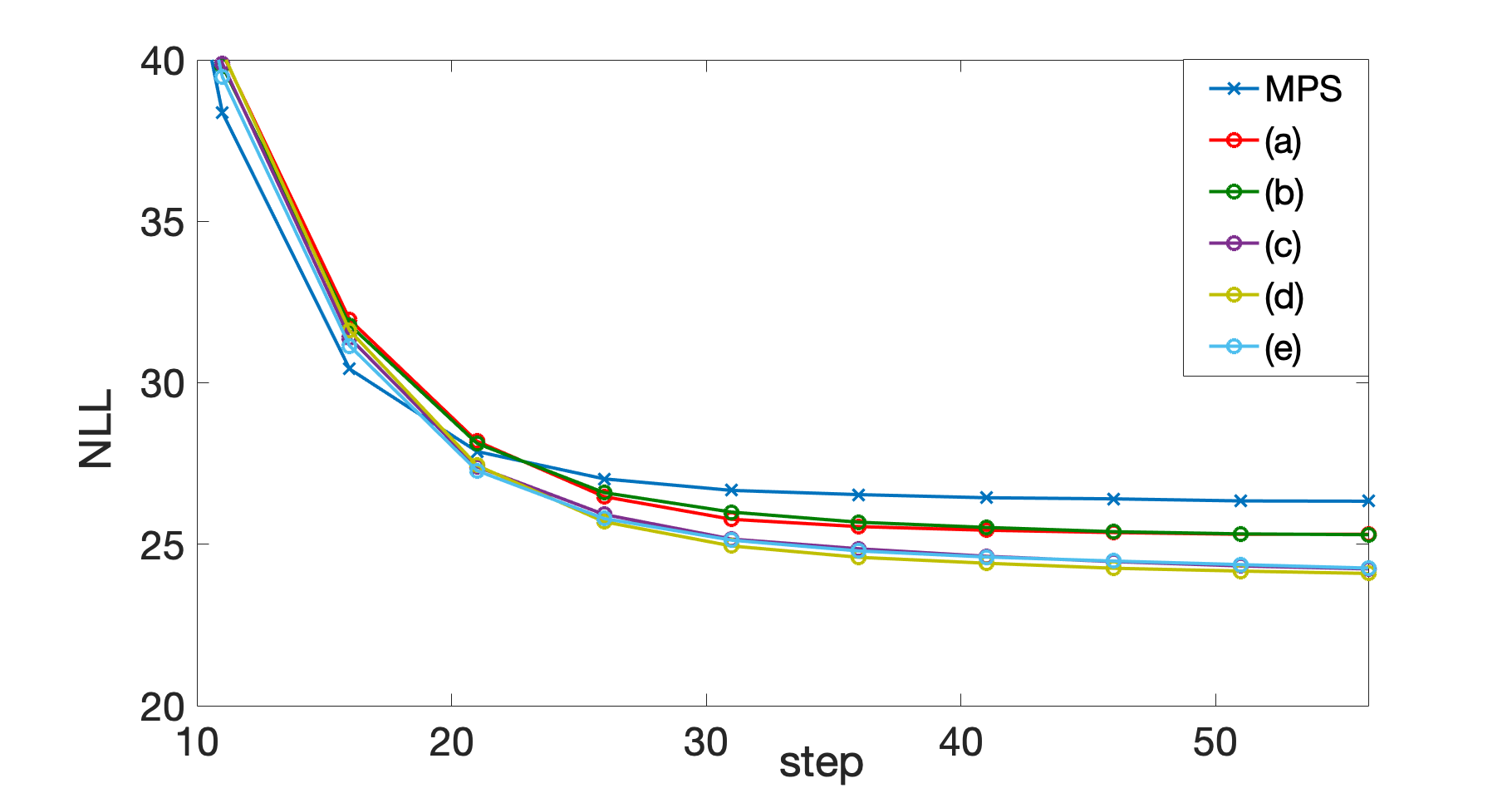}
    \caption{Comparison of performance for MPS and SMPSes in  generative modeling. The label (a)-(e) are correspond with FIG.
\ref{fig:stt_mnist}}
    \label{fig:nll}
\end{figure}

Now we can gain some insights on where to add the shortcut, based on the correlations generated by MPS. 
First we do not want to add shortcut to nearby spins, because the correlations there are well captures by the MPS (like (b) in Fig.~\ref{fig:stt_mnist} and Fig.~\ref{fig:nll}). 
We also do not want to add shortcut to the black areas (such as point (a) in Fig.~\ref{fig:stt_mnist} and \ref{fig:nll}, for example, the top and bottom of the figure, in which places there are very few range correlations to preserve.
The best locations for the shortcuts can be observed by comparing with Fig.~\ref{fig:tt_mnist}. The correlation locates in the top right corner and bottom left corner, which represent the long distance interactions, are difficult for MPS to store. 
For verifying the efficiency of the choice of shortcut and SMPS with $d_{max}=20$, we train the SMPS using the same $50$ images from the MNIST dataset, with different shortcut locations , the result are shown in Fig.~\ref{fig:stt_mnist} and Fig.~\ref{fig:nll}.

\begin{figure}[]
    \centering
    \includegraphics[width=0.8\textwidth]{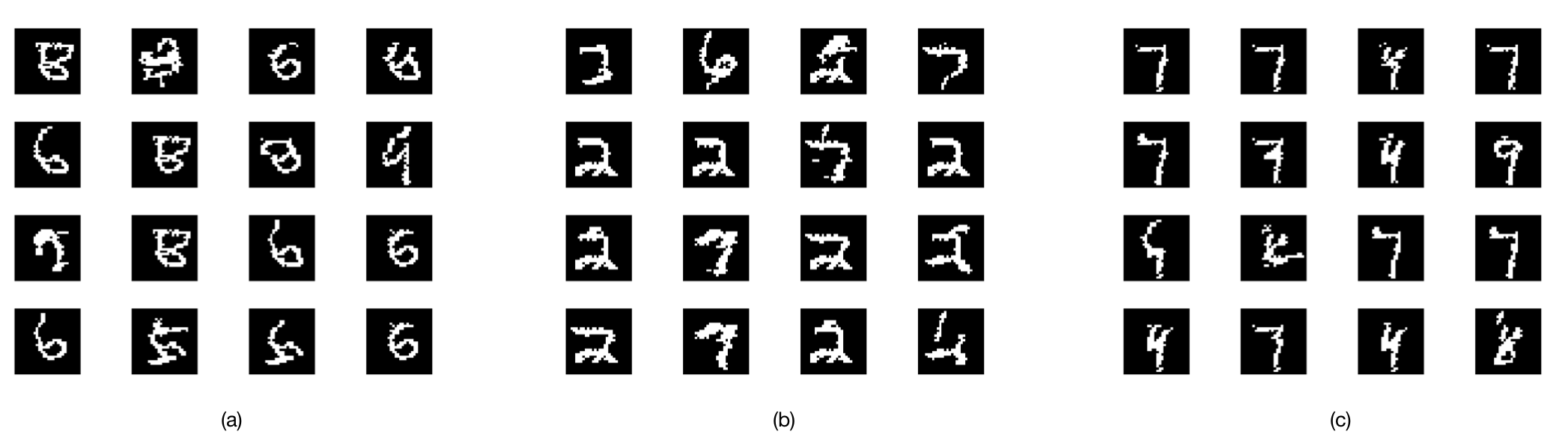}
    \caption{Images generated from different sub-MPSes.}
    \label{fig:generated}
\end{figure}

We can see from the figure that the shortcuts increase the all interactions among the linked tensors improve the performance over MPS.
For case (a) and (b) where the shortcut was added on the nearby tensors, the effect of increasing correlation span is not obvious, as we have just discussed. While for other cases, we can see clearly that the shortcut induces correlations. 
As we established in Theorem ~\ref{sum},  SMPS can be viewed as the summation of several MPSes, some internal structures of the training set can be demonstrated from these sub-MPSes. We randomly choose $3$ sub-MPSes of a SMPS in which the $250$-th and $500$-th tensor are linked, and show several samples generated from them by using generate algorithm in~\cite{PhysRevX.8.031012}. The generated samples are plotted in Fig.~\ref{fig:generated}). It is clear to see from the figure that each sub-MPS remembers a kind of image so that the whole SMPS can remember all the given images. 




\section{Conclusions and discussions}
We have presented shortcut matrix product states as a variant of MPS for preserving long range correlations. The correlation function are used in predicting the best locations for the shortcut. We derive decomposition algorithms for SMPS from a raw tensor based on a sequence of singular value decompositions, as well as learning algorithms using gradient descent to decrease loss functions specific for applications. Some rigorous properties of SMPS are also established. 

We have applied SMPS to several applications, and show that 
in the function and image fitting the shortcut in the SMPS improves the performance over MPS in tensor representation and reveals the internal structures of the data; in the application of computing partition function of Ising model, where the loss function is the contraction results, SMPS gives much more accurate results of free energy, energy as well as specific heat values; in the application of unsupervised generative modeling of the hand written digits, where the loss function is negative log-likelihood function, SMPS clearly preserves more correlations than MPS, gives a lower negative log-likelihood values.


We note here that apart from the advantage of adding shortcut for preserving long range dependences, there are limitations come as well. The most significant one is that the shortcut completely breaks the canonical conditions in MPS due to which some calculations can be simplified heavily. As an example, to calculate the normalization factor in the generative modeling, only the last tensor need to be contracted because the partition function becomes unity due to the canonical form. While with the presence of shortcut, all tensor need to be contracted explicitly to compute the partition function. The other limitation is that the shortcut in tensor will create a huge tensor during the two site update. When updating the two linked tensors in SMPS, the size of merged tensor in update would be $2d_{max}^2\times 2d_{max}^2$. To apply SVD on this matrix would be time consuming when $d_{max}$ is large.

\begin{acknowledgments}
We thank Lei Wang, Song Cheng, Jing Chen and Zhiyuan Xie for helpful discussions.
P.Z. is supported by Key Research Program of Frontier Sciences, CAS, Grant No. QYZDB-SSW-SYS032 and Project 11747601 of National Natural Science Foundation of China.
Part of the computation was carried out at the High Performance Computational Cluster of ITP, CAS.
\end{acknowledgments}


%

\section{Appendix}\label{sec:apx}
\subsection{Bound for change of log likelihood after performing SVD and spectrum truncation during the two-site update}
In this section we establish precisely how SVD with spectrum truncation change the NLL values during the two-site update learning introduced in Sec.~\ref{sec:loss} and in \cite{PhysRevX.8.031012}.
We demonstrate the ability of the SVD in keeping the negative log-likelihood stable during the update.
Let us use $A$ to denote the original tensor obtained by merging two $3$-way tensors,, and $B$ to denote the tensor after SVD and spectrum truncation performed on $A$. Then $T_A$ is used to denote the orginal MPS with $A$, and $T_B$ is the tensor with $B$, i.e. after SVD and truncation. 

We assume $L_n(B)<\infty$, we have:
\begin{equation}
    \begin{split}
        &L_n(T_{A})=L_n(T_{B})+\frac{\partial L_n(T_{B})}{\partial B}\left(A-B\right)\\
       &L_n(T_{A})-L_n(T_{B})=2 \left[ \frac{\partial Z}{\partial B}-\sum_\tau\frac{\partial \Psi}{\Psi \partial B} \right]\left(A-B\right)
    \end{split}
\end{equation}
It is easy to verify that:
\begin{equation}
    \begin{split}
        \left[ \frac{\partial Z}{\partial {B}}-\sum_\tau\frac{\partial \Psi}{\Psi \partial B} \right]B=0
    \end{split}
\end{equation}
Matrix $B$ is from the result of the SVD of matrix $A$:
\begin{equation}
    \begin{split}
        &A=USV'\\
    \end{split}
\end{equation}
During the Reconstruction of bond, only the first $d$ singular values are kept, and the partition function is kept to 1, then we have:
\begin{equation}
\begin{split}
&S^{\star}_{ii}=\left\{
             \begin{array}{lr}
            S_{ii}, i\leq d\\
            0, i>d
             \end{array}
\right.\\
&\|S^{\star}\|=1-\varepsilon\\
&B=\frac{U}{1-\varepsilon}S^{\star}V'
\end{split}
\end{equation}
Let $D=U'\left[ \frac{\partial Z}{\partial {B}}-\sum_\tau\frac{\partial \Psi}{\Psi \partial {B}} \right]V'$, we get:
\begin{equation}
    \begin{split}
        L_n(T_{A})-L_n(T_{B})&=2 \left[ \frac{\partial Z}{\partial {B}}-\sum_\tau\frac{\partial \Psi}{\Psi \partial {B}} \right]\left(B-A\right)
        =\mathrm{Tr}(2DS)=\sum_{i=1}^{}2D_{ii}S_{ii}\\
        &=2\sum_{i=d+1}D_{ii}S_{ii}
        \leq 2\sqrt{(\sum_{i=d+1}D_{ii}^2)(\sum_{i=d+1}S_{ii}^2)}\\
        &= 2\sqrt{(\sum_{i=d+1}D_{ii}^2)(2\varepsilon-\varepsilon^2)}\leq 2\sqrt{(\sum_{i,j}D_{ij}^2)(2\varepsilon-\varepsilon^2)}\\
        &\leq \sqrt{8\varepsilon}\|D\|_2=\sqrt{2\varepsilon}\left\|\frac{\partial L_n(A)}{\partial {B}}\right\|_2
    \end{split}
\end{equation}

\end{document}